\documentclass[12pt]{article}
\usepackage{latexsym}
\usepackage{epsfig}

\usepackage{graphicx}
\usepackage{epstopdf}

\usepackage{epsfig,amssymb,amsmath,amscd}

\newcommand{\bq}{\begin{equation}}
\newcommand{\eq}{\end{equation}}
\newcommand{\bqa}{\begin{eqnarray}}
\newcommand{\eqa}{\end{eqnarray}}
\newcommand{\ben}{\begin{enumerate}}
\newcommand{\een}{\end{enumerate}}
\newcommand{\bc}{\begin{center}}
\newcommand{\ec}{\end{center}}
\newcommand{\bqb}{\begin{eqnarray*}}
\newcommand{\eqb}{\end{eqnarray*}}

\def\pr#1#2#3{Phys. Rev. ${\bf{#1}}$, #2 (#3)}

\def\pl#1#2#3{Phys. Lett. ${\bf{#1}}$, #2 (#3)}

\def\np#1#2#3{Nucl. Phys. ${\bf{#1}}$, #2 (#3)}

\def\jhep#1#2#3{JHEP ${\bf{#1}}$, #2 (#3)}
\def\epj#1#2#3{Eur. Phys. J. ${\bf{#1}}$, #2 (#3)}

\def\jmp#1#2#3{J. Mod. Phys. ${\bf{#1}}$, #2 (#3)}

\begin{document}
\pagenumbering{arabic}
\thispagestyle{empty}
\def\thefootnote{\fnsymbol{footnote}}
\setcounter{footnote}{1}

\begin{flushright}
Dec. 14, 2017\\
 \end{flushright}

\begin{center}
{\Large {\bf Possible search for dark matter connected to 
massive standard particles in $e^+e^-$ collisions}}.\\
 \vspace{1cm}
{\large F.M. Renard}\\
\vspace{0.2cm}
Laboratoire Univers et Particules de Montpellier,
UMR 5299\\
Universit\'{e} Montpellier II, Place Eug\`{e}ne Bataillon CC072\\
 F-34095 Montpellier Cedex 5, France.\\
\end{center}

\vspace*{1.cm}
\begin{center}
{\bf Abstract}
\end{center}

We assume that the mass of the heavy standard particles ($Z,W,t,...$)
arises from a special coupling with dark matter and that this implies 
a corresponding peculiar connection of these particles to the dark sector.
We give examples of possible tests of this assumption in $e^+e^-\to Z+X,~
W^+W^-+X, ~t\bar t+X$ and we mention other possible studies.

\vspace{0.5cm}

\def\thefootnote{\arabic{footnote}}
\setcounter{footnote}{0}
\clearpage

\section{INTRODUCTION}

Dark matter (DM) has been discovered through its 
gravitational interaction with usual matter. DM particles have been
and are still searched in various processes at the high energy colliders. 
For a recent review with many references see \cite{rev1}.\\
Many (hundreds of) models have been proposed for the structure of dark matter and
its possible (necessary weak) interaction with usual matter through 
various types of extensions of the standard model (SM). 
No signal has been found yet. This may be due to the very small strength 
of the involved interactions or to the very heavy mass of the DM constituents.\\
Probably because of its invisibility DM should not have gauge (strong, weak, em) 
interactions with SM particles, but only some new type of interaction.\\ 
In the worst case DM may consist of a new set of particles with their own 
self-interactions and only gravitational interactions will appear 
with the SM sector.\\
But the role played by the mass may suggest other possibilities.\\
One of them consists in assuming that, both in the SM sector and in the dark sector,
mass is generated by a Higgs mecanism. If this mechanism is common to the two
sectors then it may generate a connection between them; for example simply through
Higgs boson exchange or even through a richer set of Higgs bosons; 
see \cite{Portal} for the "portal" concept.\\
Another possibility is that the mass of the SM particles is generated
through special interactions with the dark sector.
A naive picture consists in assuming that SM particles get their mass
from an environment of DM. This may be the origin of the Higgs potential
generating the SM masses but also additional interactions between SM and DM particles.\\
We propose tests of these two possibilities with processes involving the
heaviest SM particles ($Z$, $W$, top quark,...) which should be
the best places for revealing a connection with DM,  
either a direct SM-DM connection or a connection
through Higgs boson (and possibly other bosons) exchanges betwen SM and DM.\\
We are not working with a precise model for the connection to DM; we
use a kind of effective coupling between heavy SM particles and invisible
DM ones.
For example in the case of the $Z$ boson no gauge coupling "Z-DM-DM" would
exist but only some type of mass generating coupling "Z-Z-DM" would appear.
A similar assumption will be done for $W$ and for the top quark.\\
In this paper we consider the $e^+e^-$ collision processes
$e^+e^-\to Z+X,~W^+W^-+X, ~t\bar t+X$ where $X$ represents the invisible
dark matter.\\
Using the mentioned effective couplings we compute the
inclusive mass distributions ${d\sigma\over dM_X}$ and discuss their shape
reflecting the type of connection between the heavy SM particle and the
invisible DM ones.\\
We conclude by mentioning that other processes (more difficult to analyze)
may be considered for example in hadronic or in photon-photon collisions.\\

\section{Analysis of $e^+e^-\to Z+X$}

As explained in the introduction one possibility of connection
to invisible matter would be through the mass generation of the $Z$ boson.\\
Without a specific model the description can be done with an effective
$Z-Z-DM$ coupling.\\
With a kind of substructure picture for the dynamical generation of the mass 
(like in the hadronic case with quark binding) we will use a $ZZss'$ coupling
with a pair of scalars $s,s'$ DM "partons" which then (like hadronic partons) automatically 
create the multiparticle DM final (invisible) state.
The $Z$ mass may be generated according to the picture of the upper diagram
in Fig.1. The corresponding DM emission will appear from the left lower level
diagram.\\

One may then also consider (as a part of the above contribution or as an additional one)  
the possibility of Higgs boson production
like in the SM case with the process  $e^+e^-\to Z \to Z+H$ then
completed by the $H \to DM$ vertex; see the right lower level diagram of Fig.1. 
This would correspond to the idea that the Higgs boson,
and possibly heavier (excited?) $H'$, are portals to the dark sector
\cite{Portal}.\\

Several previous works about this process have been done, for example 
in \cite{ZH} for
the search of possible signals of Higgs boson compositeness  
\cite{Hcomp2,Hcomp3,Hcomp4}.
The possibilities at high energy $e^+e^-$ colliders have been reviewed
in \cite{Moortgat, Craig}.\\

We have computed the inclusive (invisible) $M_X$ mass distributions 
${d\sigma\over dM_X}$ associated to these different possibilities
corresponding to the diagrams of Fig.1 where
$M^2_X=(p_{e^+}+p_{e^-}-p_Z)^2$. \\
In Fig.2 we have plotted them (as well as their total) by using 
arbitrary couplings and masses in order that the shapes appear clearly,
with $m_{H'}=0.7$ TeV, $\Gamma_{H'}=0.1$ TeV and 
(0.01, 0.1, 0.5 TeV) for the $s,s'$ masses.\\ 
In the upper level we show the individual $H$, $H'$ peaks and the 
continuum due to the effective $ZZss'$ coupling for a parton mass of $0.01$ TeV.
In the lower level we show the total for the three choices of parton mass
(0.01, 0.1, 0.5 TeV) with clear threshold effects.\\
The quantitative values have no predictive meaning, our aim is just to see
what effects these various dynamical assumptions can produce.\\
Precise analyses should also consider backgrounds with invisible productions
like $e^+e^-\to ZZ$ followed by one $Z\to\nu\bar\nu$ decay producing a peak
at $M_X=M_Z$.\\

\section{Analysis of $e^+e^- \to W^+W^-+X$}

A similar analysis can be done for the process $e^+e^- \to W^+W^-+X$.\\
We had previously considered this process among others for the search of signals
of $W$ compositeness, see \cite{CSMrev}. Now we look more specifically at
the hypothesis that DM is related to mass generation
of the $W$ boson, with the same procedure as above for the $Z$ boson, 
and that this can be described with an effective 
coupling to an $ss'$ pair of subconstituents,  $W^+W^- ss'$.\\
We then compute the corresponding inclusive (invisible) $M_X$ mass distributions 
${d\sigma\over dM_X}$ according to the diagrams of Fig.3.\\
Note that, at the same order, a contribution still appears through the  
$Z-Z-DM$ coupling.\\
As above we add the possibility of intermediate exchange of Higgs bosons
$H$ and $H'$connected to DM.\\
We make the illustrations with the same parameters as in the previous $Z$
process.\\
Fig.4 shows that effects similar to those observable in $e^+e^-\to Z+X$
could confirm the basic hypothesis. Note also the presence of the background
$e^+e^- \to W^+W^-Z$ with $Z\to\nu\bar\nu$.\\

\section{Analysis of $e^+e^- \to t\bar t+X$}

Finally we consider the process $e^+e^- \to t\bar t+X$ that has also been a part
of the studies, for example \cite{Tait, trcomp, ttincl}, 
of top quark compositeness \cite{partialcomp}. Substructure models have been
proposed since a long time \cite{comp}.\\
We again assume that DM is related to mass generation
of the top quark and that this can be described with an effective 
coupling to an $ss'$ pair of subconstituents, $t\bar t ss'$.\\
We then compute the corresponding inclusive (invisible) $M_X$ mass distributions 
${d\sigma\over dM_X}$ according to the diagrams of Fig.5.\\
Like in the $e^+e^- \to W^+W^-+X$ case a contribution involving the  
$Z-Z-DM$ coupling also appears.\\
Diagrams with intermediate exchange of Higgs bosons
$H$ and $H'$ are also added.\\
The background from $e^+e^- \to t\bar tZ$ with $Z\to\nu\bar\nu$ is also present.\\
In Fig.6, with the same parameters as in the previous
processes, we can see that it should be possible to check if the
hypothesis of DM connection related to mass generation of the Z,W gauge bosons 
also applies to the top quark.\\

\section{Conclusion and further developments}

In this paper we have assumed that invisible dark matter,
in addition to gravitational interaction, may have other
types of interactions with standard particles related to
their mass.\\
This assumption suggests studies of $Z$, $W$ and the top quark production
for revealing this property.\\
Using arbitrary effective couplings describing this special interaction we
have computed the invisible inclusive distribution ${d\sigma\over dM_X}$
for the three processes $e^+e^-\to Z+X,~W^+W^-+X, ~t\bar t+X$. 
The illustrations show how the shapes of these distributions reflect
the various possibilities and the parameters controlling the effective
couplings to DM.\\
Other processes involving heavy SM particles and DM may be considered.\\
In $e^+e^-$ collisions, see \cite{Moortgat} for a general review, 
one example is $e^+e^-\to e^+e^-+ZZ$ with
$ZZ$ fusion into DM (directly or through Higgs-like bosons).\\
 
Obviously many other processes involving $Z$, $W$ and the top quark
occur in hadronic collisions but require detailed and difficult
phenomenological and experimental analyses, see \cite{rev1} and also
\cite{Contino,Richard}.\\

Studies in photon-photon collisions may also be considered \cite{gammagamma}.\\

Other less massive SM particles could also be concerned. It is not yet experimentally
established that the Higgs couplings  proportional to the fermion masses. 
So non negligible direct couplings to dark matter may even also appear there.
This may concern $b$ quark, muon and even electron. If a direct $e^+e^-$ 
coupling is too small (see although \cite{light}), 
a muon collider, already known as a possible Higgs boson factory, 
could be an interesting place.  One would look
at the process $\mu^+\mu^- \to \gamma +DM$, involving
$\mu^+\mu^- \to DM$ direct production or 
$\mu^+\mu^- \to H,H'\to DM$ and a photon emission from the $\mu^{\pm}$ line.
Obviously this has to overwhelm (at least in some $M_X$ range)
the background involving $Z\to\nu\bar\nu$.\\

Finally we should obviously add that quantitative
predictions require the use of a precise theoretical description
of DM and of its possible relation to the mass generation.\\

\newpage
\begin{figure}[p]

\[
\epsfig{file=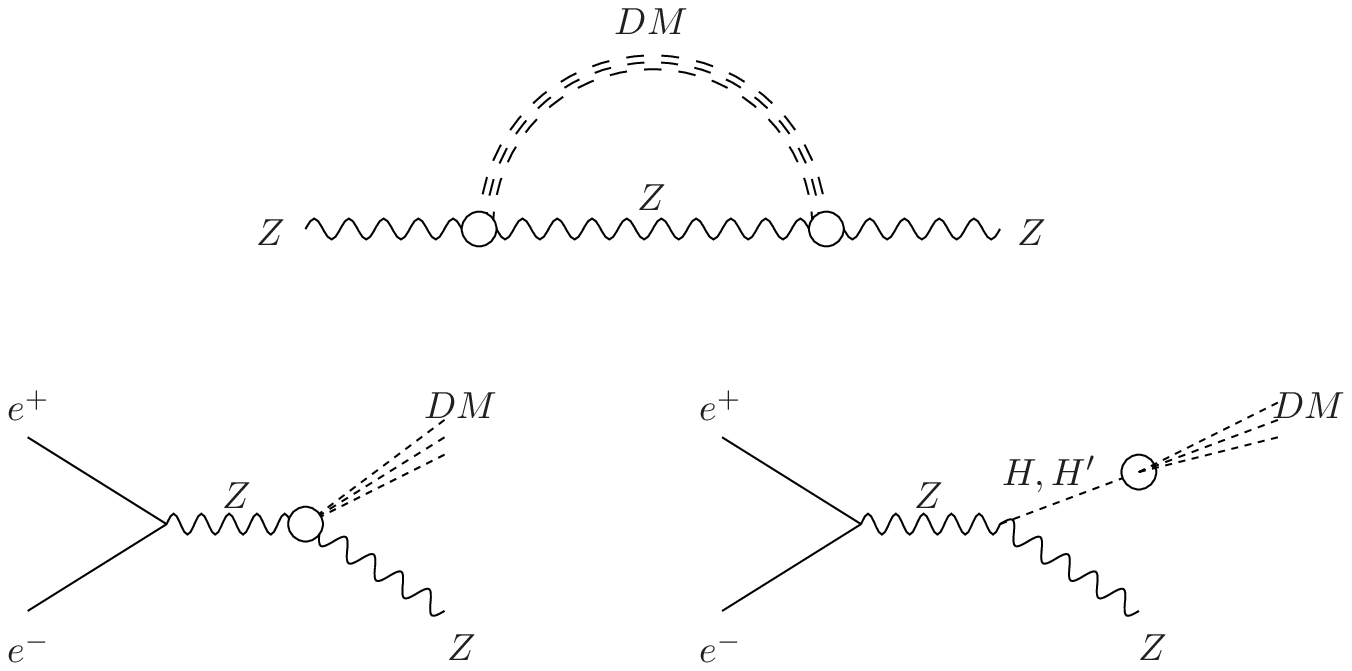 , height=8.cm}
\]\\
\caption[1] {Diagrams for $Z$ mass generation (upper level)
and for $e^+e^-\to Z+X$ (lower level), direct continuum production (left diagram)
and through $H$ and $H'$ (right diagram).}
\end{figure}

\clearpage

\begin{figure}[p]
\[
\epsfig{file=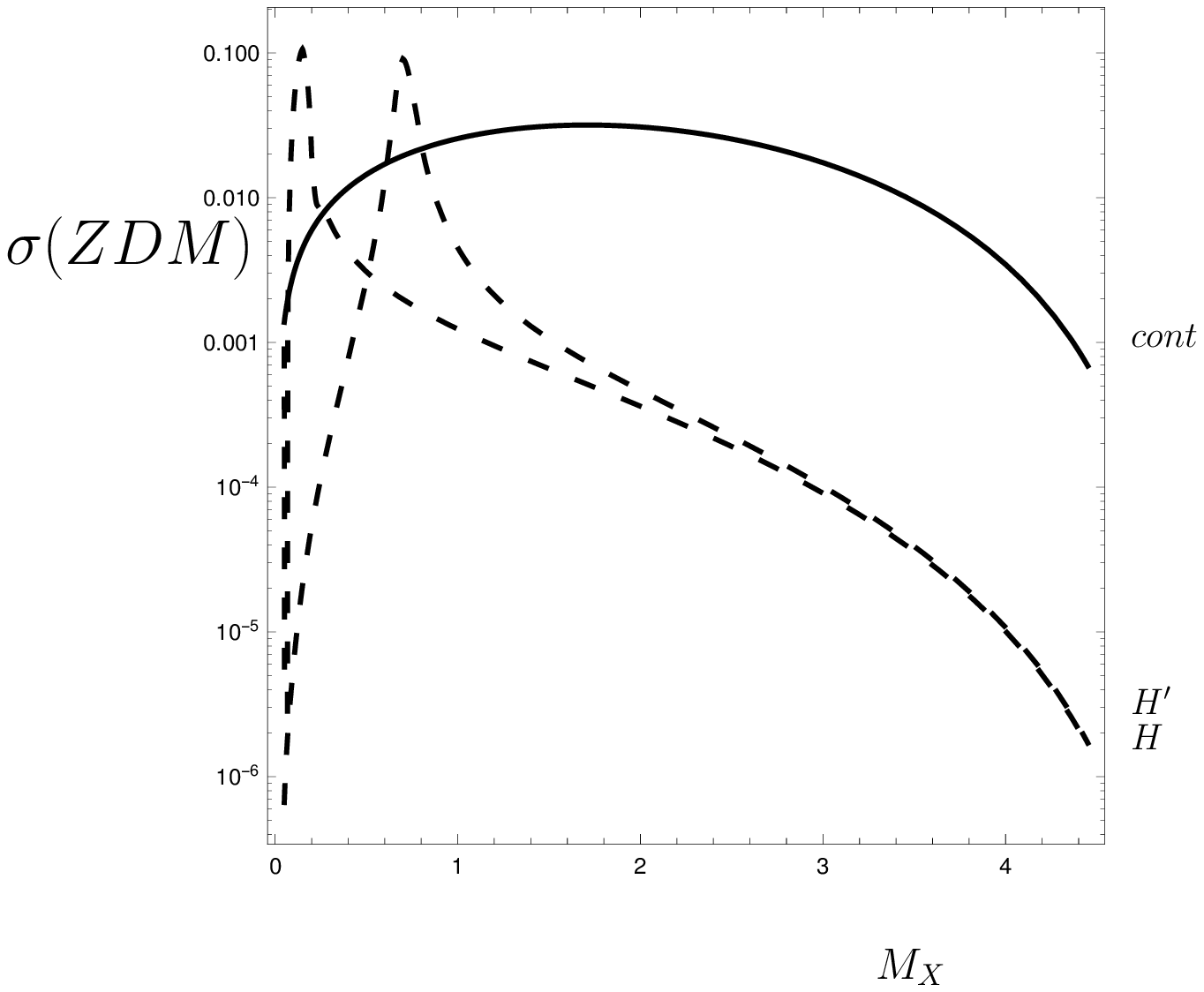 , height=8.cm}
\]\\
\[
\epsfig{file=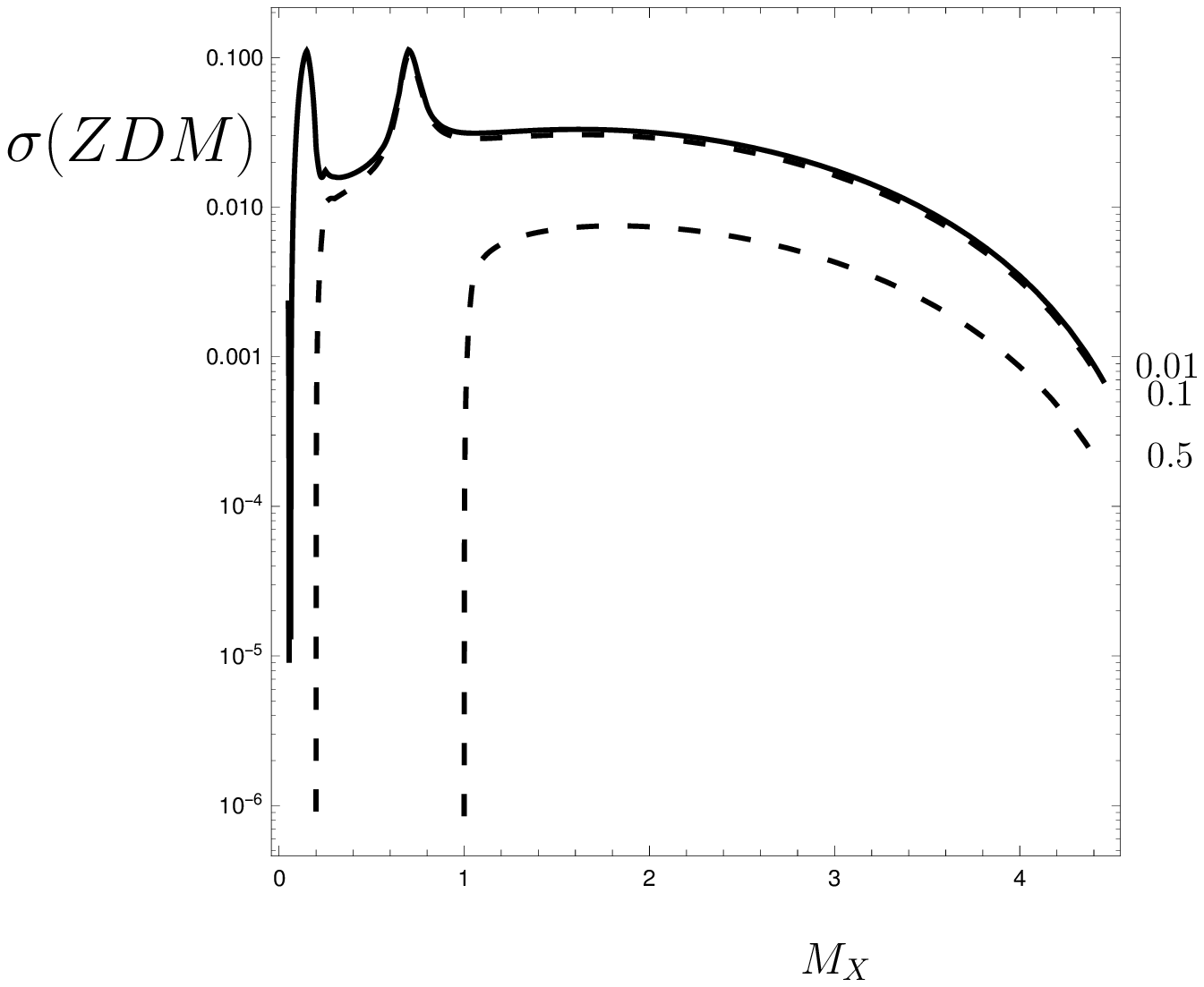 , height=8.cm}
\]\\

\caption[1] {DM production in $e^+e^-\to Z+X$; upper level
for $m_s=0.01$ TeV:
direct continuum, through H, through H'; lower level:
total for $m_s=0.01,0.1,0.5$ TeV.}
\end{figure}

\clearpage

\begin{figure}[p]
\[
\epsfig{file=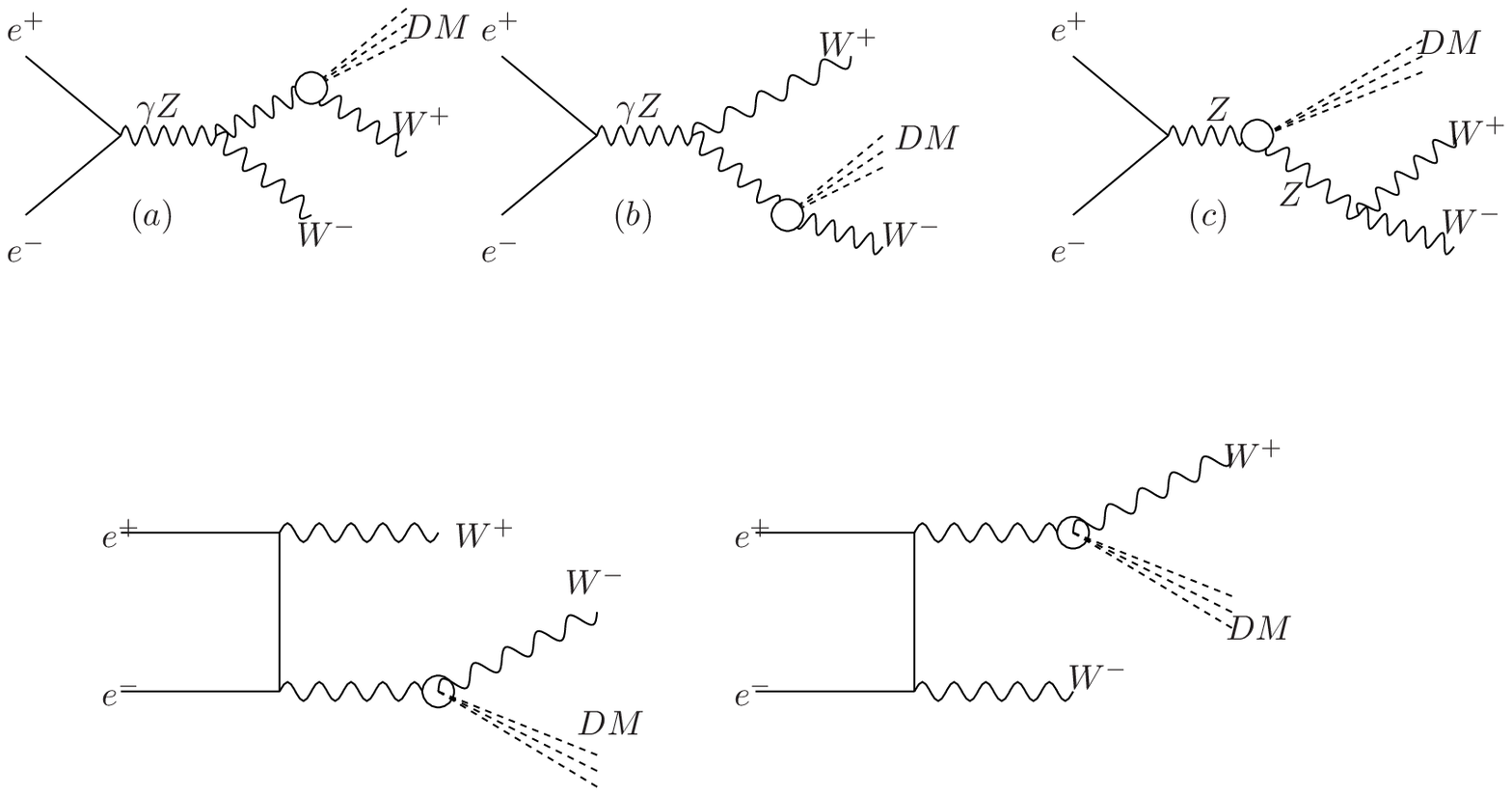 , height=8.cm}
\]\\

\caption[1] {Diagrams for  $e^+e^- \to W^++W^-+X$; direct continuum production from $W$ and $Z$ masses;
production through $H$ and $H'$ can be deduced in analogy with the $Z$ case of Fig.1.}
\end{figure}

\clearpage

\begin{figure}[p]
\[
\epsfig{file=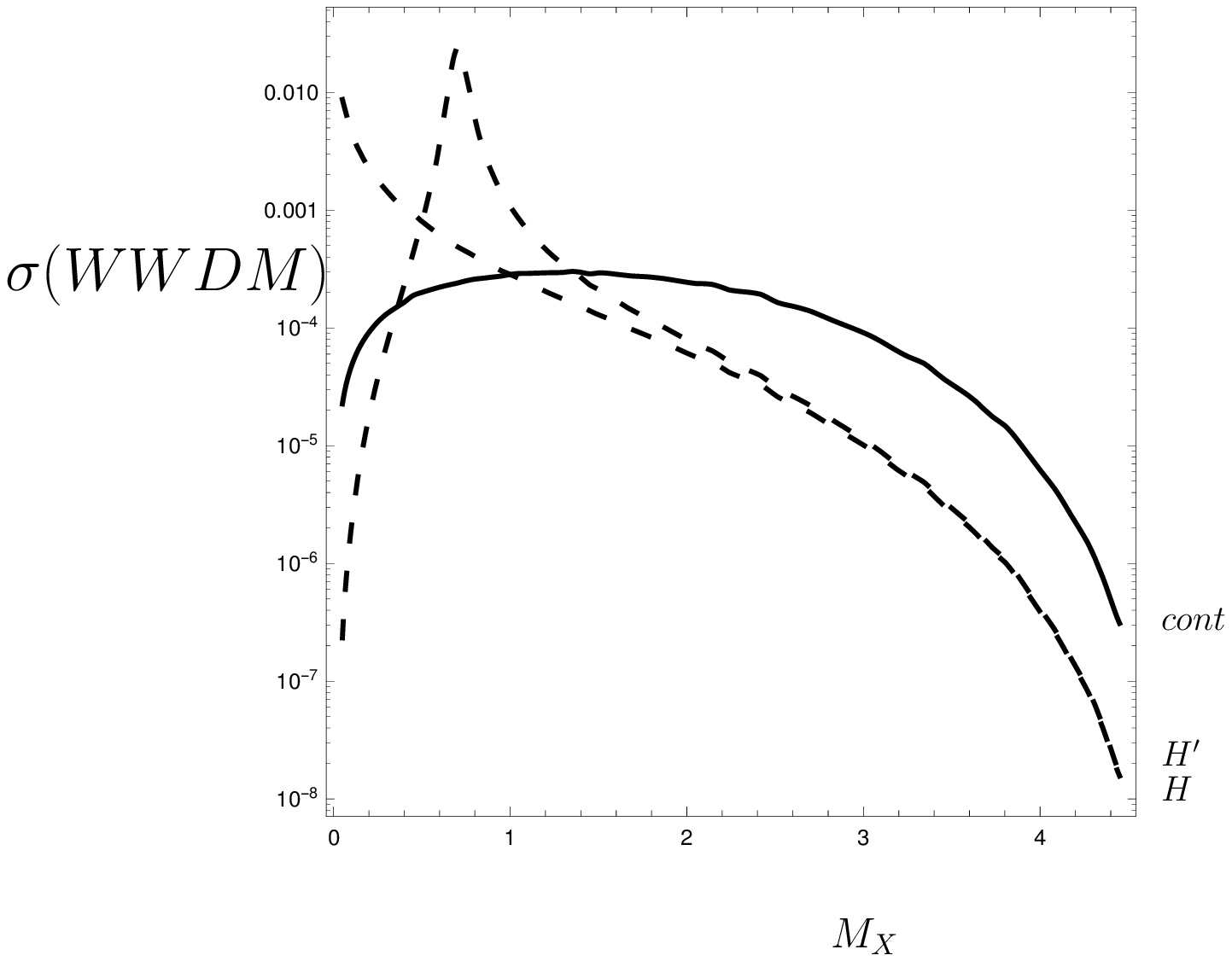 , height=8.cm}
\]\\
\[
\epsfig{file=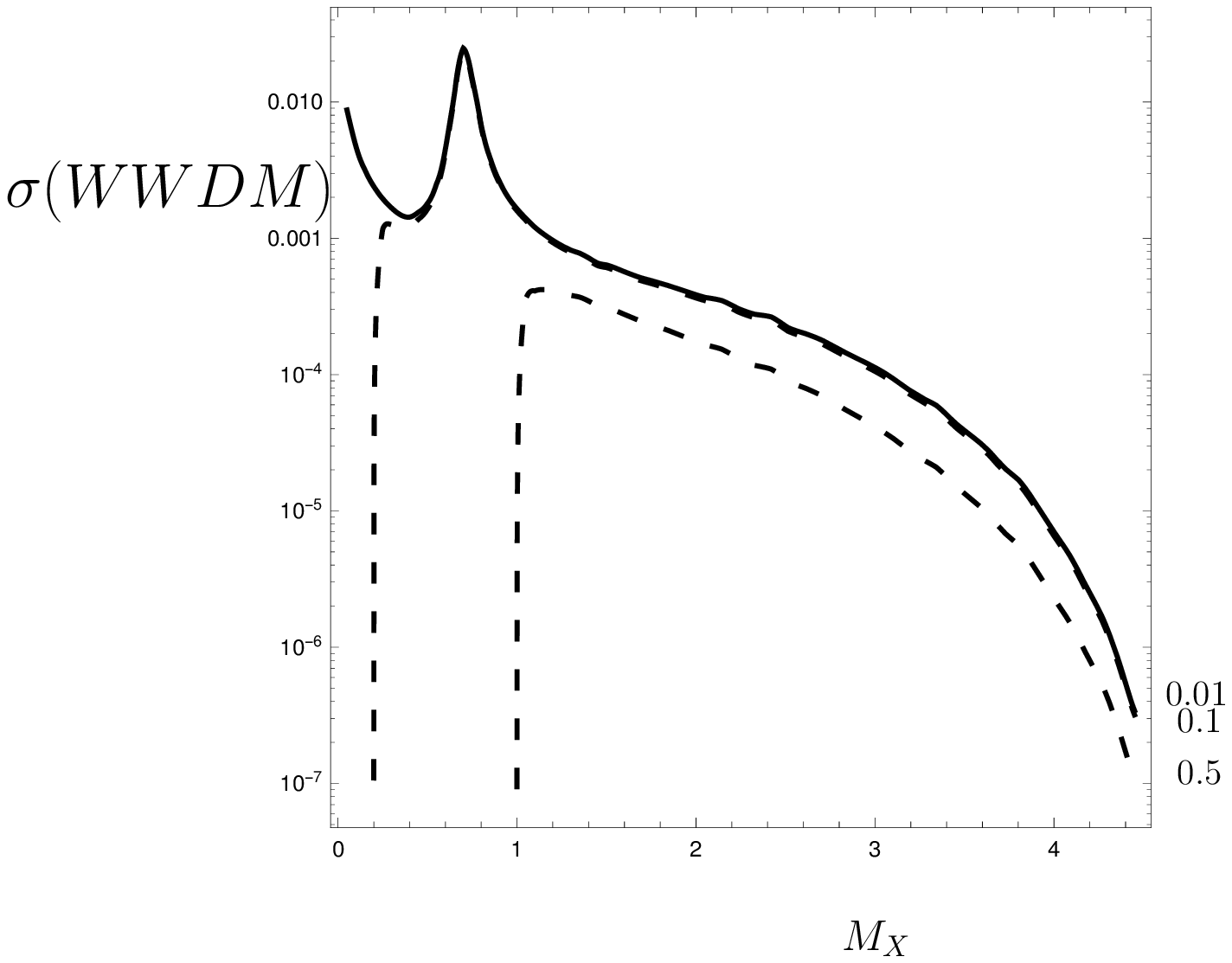 , height=8.cm}
\]\\

\caption[1] {DM production in $e^+e^- \to W^++W^-+X$; upper level
for $m_s=0.01$ TeV:
direct continuum, through H, through H'; lower level:
total for $m_s=0.01,0.1,0.5$ TeV.}
\end{figure}

\begin{figure}[p]
\[
\epsfig{file=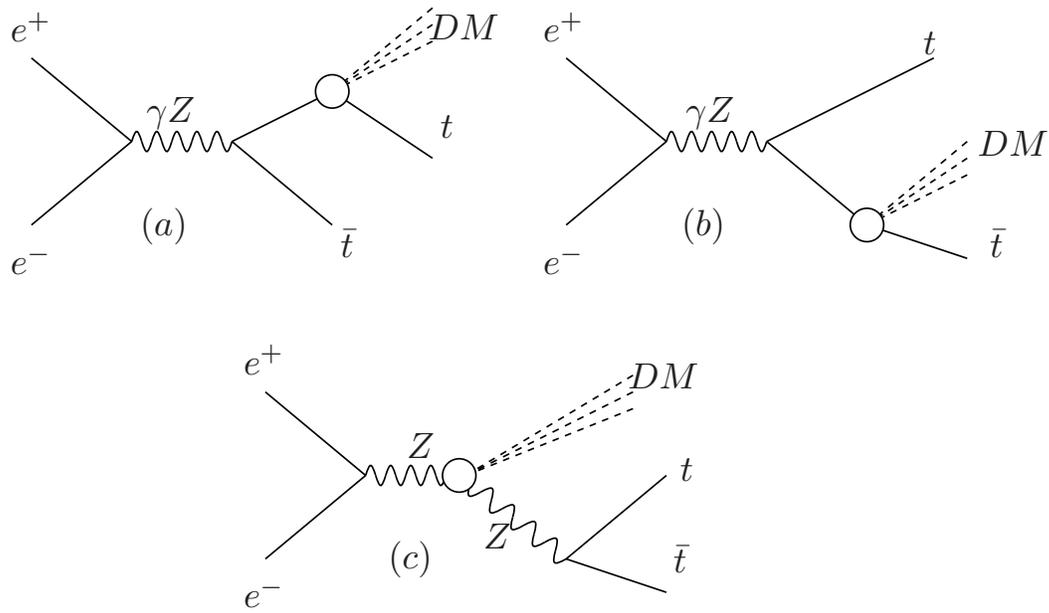 , height=8.cm}
\]\\

\caption[1]  {Diagrams for $e^+e^-\to t\bar t+X$; direct continuum production from top quark and $Z$ masses; production through $H$ and $H'$ can be deduced in analogy with the $Z$ case of Fig.1.}
\end{figure}

\clearpage

\begin{figure}[p]
\[
\epsfig{file=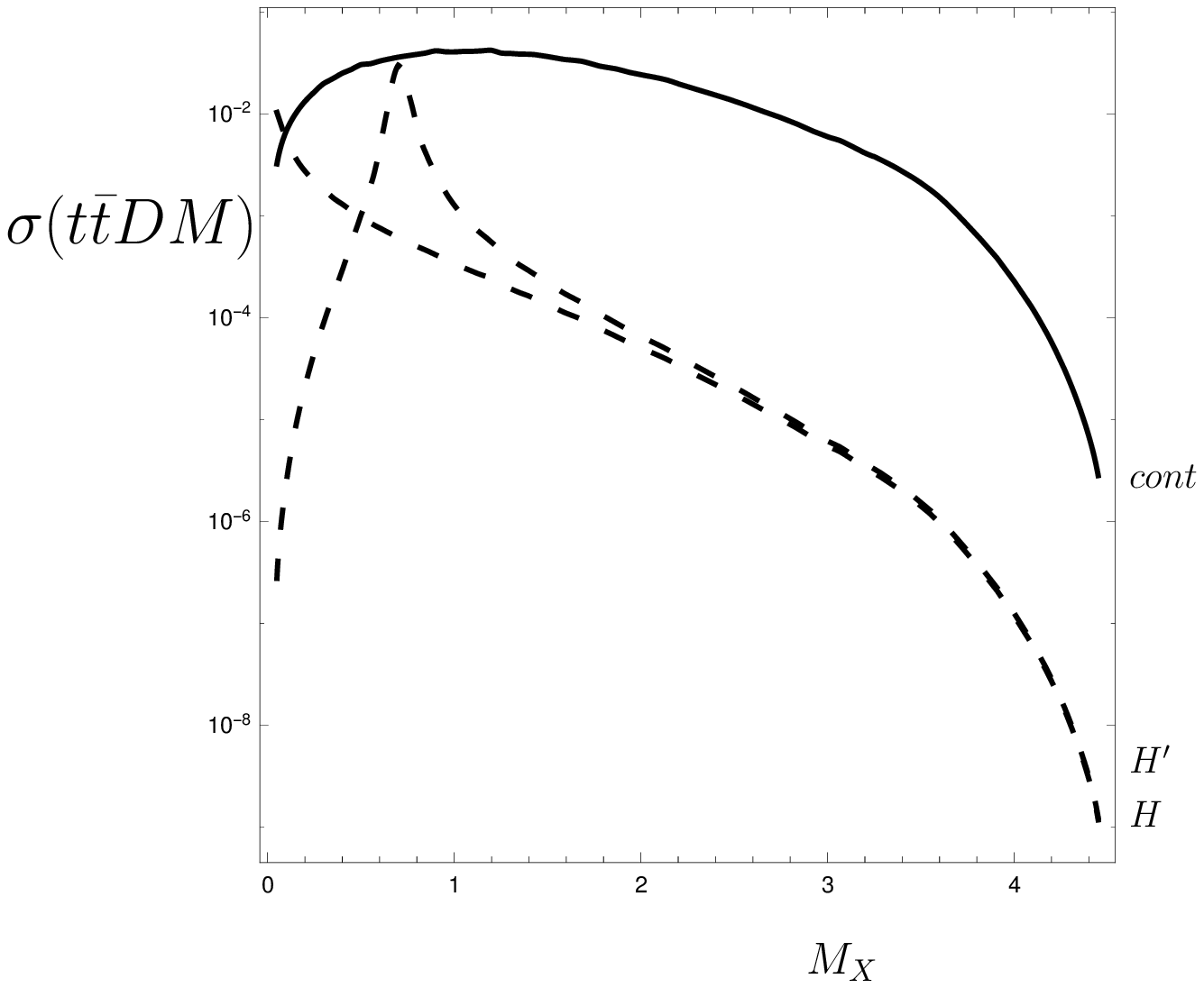 , height=8.cm}
\]\\
\[
\epsfig{file=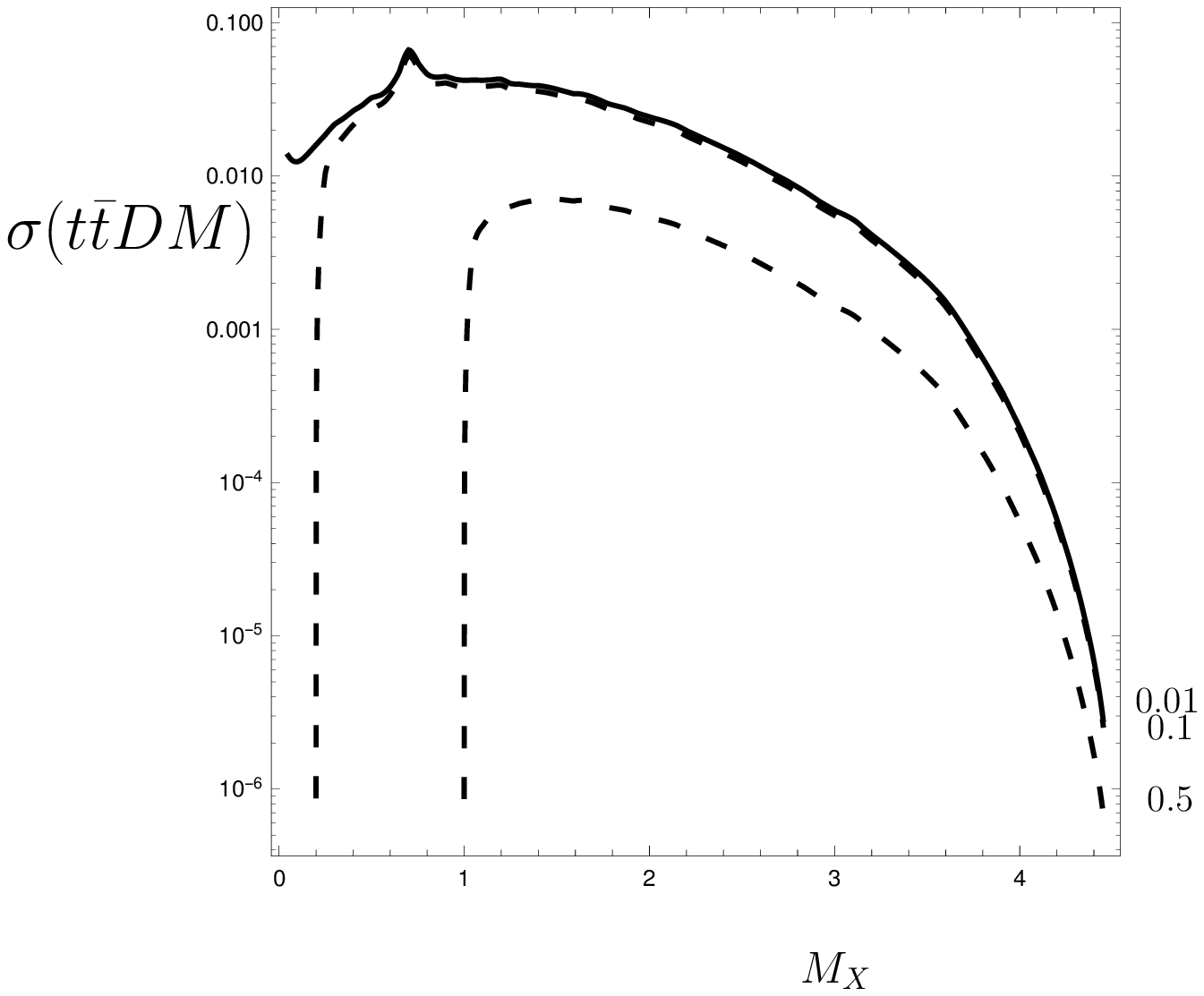 , height=8.cm}
\]\\

\caption[1] {DM production in $e^+e^- \to t+\bar t+X$; upper level
for $m_s=0.01$ TeV:
direct continuum, through H, through H'; lower level:
total for $m_s=0.01,0.1,0.5$ TeV.}
\end{figure}

\clearpage


\begin{thebibliography}{99}

%
\bibitem{rev1} B. Penning, arXiv: 1712.01391.
%
\bibitem{Portal} B. Patt and F. Wilczek, arXiv: 0605188 (hep-ph].

%
\bibitem{ZH}  F.M. Renard, arXiv: 1701.04571.
%
\bibitem{Hcomp2} D.B. Kaplan and H. Georgi, \pl{136B}{183}{1984}.
%
\bibitem{Hcomp3} K. Agashe, R. Contino and A. Pomarol, \np{B719}{165}{2005}; hep/ph 0412089.
%
\bibitem{Hcomp4} G. Panico and A. Wulzer, Lect.Notes Phys. {\bf 913},1(2016).
%
\bibitem{Moortgat} G.  Moortgat-Pick et al, \epj{C75}{371}{2015}, arXiv: 1504.01726.
%
\bibitem{Craig} N. Craig, arXiv: 1703.06079.

\bibitem{CSMrev}  F.M. Renard, arXiv: 1708.01111. 
%
\bibitem{Tait} Ben Lillie, Jing Shu, Timothy M.P. Tait,  \jhep{0804}{087}{2008}, 
arXiv:0712.3057; Kunal Kumar, Tim M.P. Tait, Roberto Vega-Morales
\jhep{0905}{022}{2009}, arXiv:0901.3808.
%
\bibitem{trcomp}  G.J. Gounaris and F.M. Renard,
arXiv: 1611.02426.
%
\bibitem{ttincl}  F.M. Renard, arXiv: 1711.02340.
%
%
\bibitem{partialcomp} R. Contino, T. Kramer, M. Son and R. Sundrum,
J. High Energy Physics {\bf 05}(2007)074.


%
\bibitem{comp}  H. Terazawa, Y. Chikashige and K. Akama, \pr{D15}{480}{1977};
for other references see
H. Terazawa and M. Yasue, Nonlin.Phenom.Complex Syst. {\bf19},1(2016);
\jmp{5}{205}{2014}.

%
\bibitem{Contino} R. Contino et al, arXiv: 1606.09408.
%
\bibitem{Richard}  F. Richard, arXiv: 1703.05046.
%
\bibitem{gammagamma} V.I. Telnov, Nucl.Part.Phys.Proc. {\bf 273},219(2016).
%
\bibitem{light} A. Dery, C. Frugiuele and Y. Nir, arXiv: 1712.04514.


\end{thebibliography}
\end{document}